\documentclass[conference]{IEEEtran}
\usepackage{graphicx}
\usepackage{url}
\usepackage{amsmath}

\graphicspath{{images//}} 

\begin{document}

\pagestyle{plain}
\pagenumbering{arabic}

%
\title{Community Detection Using A Neighborhood Strength Driven Label Propagation Algorithm}

\author{\IEEEauthorblockN{Jierui Xie and Boleslaw K. Szymanski}
\IEEEauthorblockA{Department of Computer Science\\Rensselaer Polytechnic Institute\\
110 8th Street\\Troy, New York 12180\\Email: \{xiej2,szymansk\}@cs.rpi.edu}
}

\maketitle

\begin{abstract}
Studies of community structure and evolution in large social networks require a fast and accurate algorithm for community detection. As the size of analyzed communities grows, complexity of the community detection algorithm needs to be kept close to linear. The Label Propagation Algorithm (LPA) has the benefits of nearly-linear running time and easy implementation, thus it forms a good basis for efficient community detection methods. In this paper, we propose new update rule and label propagation criterion in LPA to improve both its computational efficiency and the quality of communities that it detects. The speed is optimized by avoiding unnecessary updates performed by the original algorithm. This change reduces significantly (by order of magnitude for large networks) the number of iterations that the algorithm executes. We also evaluate our generalization of the LPA update rule that takes into account, with varying strength, connections to the neighborhood of a node considering a new label. Experiments on computer generated networks and a wide range of social networks show that our new rule improves the quality of the detected communities compared to those found by the original LPA. 
The benefit of considering positive neighborhood strength is pronounced especially on real-world networks containing sufficiently large fraction of nodes with high clustering coefficient.

\end{abstract}


\IEEEpeerreviewmaketitle

\section{Introduction}
A network is said to have community structure if it can be divided into groups with dense connections within groups and sparse connections between groups. Community detection has received significant attention in physics and computer science communities. It has been applied to networks of many kinds, including the World Wide Web (WWW)  \cite{Hoerdt03completenessof,Broder:2000:GSW:346241.346290}, collaboration networks \cite{citeulike:128}, communication networks, social networks \cite{RefWorks:380},  biological networks \cite{Fell:2000}, and so on. This problem has been studied as the graph partitioning problem in computer science for decades and is known to be NP-hard. Algorithms that find reasonably good quality communities have been proposed \cite{Danon05comparingcommunity} and improved extensively. To name a few, they include divisive algorithms that recursively remove links from the network \cite{GirvanNewman:2002}, agglomerative algorithms that repeatedly merge smaller groups of nodes \cite{newman-2004-69,PhysRevE.69.026113}, maximization of modularity algorithms using spectral clustering, simulated annealing or extremal optimization \cite{PhysRevE.70.066111,springerlink:10.1140/epjb/e2004-00125-x,PhysRevE.74.036104,Duch:2005} and so on. The quality of detected communities can be measured using \textit{modularity} defined by Newman \cite{newman-2004-69}. A network which has its modularity in the range between 0.3 and 0.7, usually has a strong community structure. Despite great efforts, the cost of detecting \textit{unknown number} of communities of \textit{unknown size} in an arbitrary complex network remains high. For large networks, algorithms with complexity of 
$O(n^2)$, where $n$ denotes the number of nodes in the network, become prohibitively expensive in terms of their execution time. 

Recently, Raghavan \cite{Raghavan:2007} proposed a method called \textit{label propagation algorithm} (LPA) to identify community in large networks, which runs linearly in the number of edges, thus linearly also in the number of nodes for sparse networks.  Initially, each node is assigned a unique label. During the iterative process, each node adopts the label in agreement with the majority of its neighbors. At the end of the algorithm, connected nodes with the same label form a community.  This algorithm provides a number of desirable qualities such as no parameters, easy implementation and fast execution for practical networks. In this paper, we empirically study and analyze a generalized update rule for LPA.

\section{Related work}
In this section, we review community detection algorithms proposed in the literature.

Girvan and Newman \cite{GirNew02} propose a divisive hierarchical clustering algorithm, referred to as GN, which consists of  four steps:  1) Calculate the betweenness for all edges in the network. 2) Remove the edge with the highest betweenness. 3) Recalculate betweenness for all edges affected by the removal.  4) Repeat from step 2 until no edges remain.  The algorithm utilizes the non-local structure information, thus it works well on real-world networks. However, a particular disadvantage of GN is that it runs in  $O(m^2n)$ on a network of $n$ nodes and $m$ edges or $O(n^3)$ on a sparse network. Newman proposed a faster algorithm, referred to as  NM, in \cite{newman-2004-69} with running time $O((m+n)n)$ or $O(n^2)$ on a sparse network. NM is an agglomerative hierarchical clustering algorithm that starts with a state in which each node is a single community. Then, it repeatedly merges pairs of communities together, choosing at each step the merger that results in the greatest increase in modularity (termed Q). In its faster version, called CNM  \cite{PhysRevE.70.066111}, the running time is reduced to $O(md \log n)$, where $d$ is the depth of the dendrogram describing the network's community structure. On a sparse network, it runs in $O(n \log ^2n)$. It is known that NM has a resolution limit, failing to find communities with sizes smaller than a certain value. A lot of work has been done to improve GN and NM. For example, some improvements attempt to strike a balance between the community size and the gain in the modularity with various refinement strategies \cite{WakitaTsurumi:2007,SchuetzCaflisch:2008}. These methods usually have complexity comparable to the original NM algorithm. 

Another fast greedy algorithm based on modularity optimization, called Louvain method is proposed in \cite{Blondel-2008}. The method consists of two phases. First, it looks for ``small'' communities by optimizing modularity locally. Second, it aggregates nodes of the same community and builds a new network whose nodes are the communities found at the previous phases. These steps are repeated iteratively until a maximum of modularity is attained. From extensive experiments, the complexity of this method scales as $O(n\log n)$ even though this has not been formally proved. Other algorithms seeking the maximization of modularity use various techniques such as spectral clustering, simulated annealing or extremal optimization \cite{PhysRevE.70.066111,springerlink:10.1140/epjb/e2004-00125-x,PhysRevE.74.036104,Duch:2005} and so on. These methods usually obtain higher values of modularity than the original NM.

Random walk has successful applications in finding community \cite{ZhouLipowsky:2006, Hu:2008, DBLP:journals/jgaa/PonsL06}. The idea behind this approach is that the walk tends to be trapped in dense parts of a network corresponding to communities. For these algorithms, the complexity of computing \textit{distance} or \textit{proximity} between all pairs of nodes exactly is $O(n^3)$. Some approximation techniques are usually used. WalkTrap (WT) proposed in \cite{DBLP:journals/jgaa/PonsL06} is built on a measure of similarity between nodes based on random walks. WT has the time complexity of $O (mn^2)$ but runs in $O(n^2 \log n)$ in most real-world cases. Markov Cluster Algorithm (MCL) proposed in \cite{StijnDongen:2000} is an unsupervised clustering algorithm based on simulations of flow. In some sense, it is a random walk with decay. By keeping only a maximum number $k$ of non-zeros elements in each column when computing the matrix multiplication, the complexity is down to $O(nk^2)$ on a sparse network. However, this algorithm is sensitive to the parameter called \textit{inflation}. 

Spectral clustering \cite{ShiMalik:2007, WhiteSmyth:2005, Capocci:2005} first embeds a network in space and then uses a fast clustering algorithm to find communities. The space is spanned by eigenvectors. The spectral optimization method proposed by Newman~\cite{PhysRevE.74.036104} runs in $O(n^2)$ on a sparse network. White and Smyth (WS) propose a fast spectral clustering algorithm in  \cite{WhiteSmyth:2005}. They reformulate the problem of modularity optimization as a discrete quadratic assignment problem. Then they relax it as a continuous one which can be solved by eigen-decomposition. Their algorithm uses the Implicitly Restarted Lanczos Method (IRLM) and k-mean and has complexity of $O(mKh+nK^2h+K^3h +nK^2t)$, where the first three terms represent complexity of IRLM while the last one represents the complexity of execution of k-mean. $m$ and $n$ denote the number of edges and nodes, $K$ stands for the maximum number of eigenvectors, $h$ is the number of iterations required for IRLM to converge, and $t$ denotes the number of iterations of k-mean algorithm. On a sparse network, the algorithm scales roughly linearly as a function of $n$.

Multi-state spin models \cite{ Blatt:1996, Reichardt:2004, randomfield:2006, Fortunato:2007, Kumpula:2007} (e.g., q-state Potts model), in which a spin is assigned to each node in a network, can also be applied to community detection. In such a setting, community detection is equivalent to minimizing the Hamiltonian of the model. The corresponding algorithms are related to the label propagation algorithms discussed below and usually are fast. However, they may require 
some prior knowledge of the networks structure (for example knowing a pair of nodes each of which belongs to a different community) in order to be able to 
apply them to community detection (e.g., Ferromagnetic Random Field Ising Model \cite{ randomfield:2006}).

The idea of propagating labels through a network has been studied by Bagrow \cite{Bagrow:2005} in his \textit{L-shell} method. Starting from a node with a label, the algorithm propagates the label step by step and includes more neighbor nodes until the end of a community is reached. The boundary of a community is identified by the threshold defined as the ratio of the number of edges inside and outside of the community. Similar idea is studied by Costa in \cite{Costa:2004}. Wu \cite{WuHuberman:2004} proposed a method which partitions a network into two communities. The network is viewed as an electric circuit, and a battery is attached to two random nodes that are supposed to be within two communities. The algorithm amounts to solving Kirchhoff equations, with two of them fixed to be 0 and 1. In other words, each node updates its value (i.e.,voltage) by taking the average of all neighbors' value. When the process converges, the voltage gap indicates the border, and two communities are identified. Although this method can be generalized to detecting multiple communities, it requires the number of communities as the input, and tends to find communities of approximately the same size.

The LPA \cite{Raghavan:2007} uses the network structure alone to guide its process and requires neither any parameters nor optimization of the objective function. It starts from a configuration where each node has a distinct label. At every step, one node (in asynchronous version) or each node (in a synchronous version) makes its own decision to change its label to the one carried by the largest number of its neighbors. By construction, as the algorithm converges, each node has more neighbors in its own community than in any of other community. One drawback of LPA is that it returns different solutions (some of them of poor quality) in different realizations. This is because the quality of LPA solution depends on the local minima it reaches. Tib\'ely and Kert\'esz \cite{TibKert:2008} show that this model is equivalent to finding the local minima of a simple Potts model \cite{Kumpula:2007}. The number of such local minima was found to be much larger than the number of nodes in the underlying network.  Barber \cite{Barber:2009} defines an equivalent objective function based on the number of edges that connect vertices with identical labels that penalize the low quality solutions. Leung \cite{Leung:2009} extends LPA by incorporating heuristics like hop attenuation score to improve the quality of the detected communities. Gregory \cite{Gregory:2010} applies the similar idea to detection of overlapping communities. Each vertex is allowed to belong to up to $v$ communities, where $v$ is the parameter of the algorithm.

In this paper, we enhance the LPA by introducing new update and label propagation rules that achieve higher speed of execution and improve the quality of the communities detected. Although the updating process in LPA can either be synchronous or asynchronous, we restrict our attention to only asynchronous version here.

\begin{table}[!t]
\centering
\caption{The number of iterations (scaled by n) required for convergence on social networks}
\label{table:compareT}
\begin{tabular}{|c|c|c|c|c|} \hline
\textbf{Network} & \textbf{n} & \textbf{org-LPA} & \textbf{speed-up-LPA} & \textbf{Ratio of imp.}\\ \hline
karate&	34&	2.78&	1.87&	1.49\\ \hline
lesmis&	77&	3.58&	1.77&	2.02\\ \hline
polbooks&	105&	4.28&	1.70&	2.52\\ \hline
football&	115&	2.78&	1.19&	2.34\\ \hline
netscience&	379&	6.12&	1.81&	3.38\\ \hline
email&	1133&	17.68&	2.54&	6.96\\ \hline
eva&	4475&	9.46&	0.45&	21.02\\ \hline
CA-GrQc&	4730&	20.56&	3.04&	6.76\\ \hline
PGP&	10680&	21.40&	1.55&	13.81\\ \hline
\end{tabular}
\end{table}

\section{Improving the Speed of LPA}
\label{sec:impspeed}

Although LPA runs in linear time, there are ways to improve the execution time in practice, and such improvements are essential when we process extremely large networks or move from offline to online detection.  The basic idea of our improvement is to avoid unnecessary updates in each iteration of the algorithm, while maintaining the overall behavior of the algorithm unchanged. As observed, at the early stage of the original LAP, most nodes are in a very diverse neighborhood. The \textit{effectiveness} of  updates (i.e., the fraction of attempted updates that result in changes to new labels) is high. However, the competition between communities is restricted only to their boundaries after a few iterations. For nodes inside any community, the updates are unnecessary, since they essentially do not change. As shown in \cite{Raghavan:2007}, after five iterations, 95\% of nodes are already correctly clustered. Additional time is required to attempt the updates that are expected to fail to change labels, so the final convergence of the algorithm is delayed. It turns out that this amount of time can be easily saved by bookkeeping the information about the boundaries of the currently existing communities. There was an initial attempt along this line of the LAP speed improvement \cite{Leung:2009}. However, unlike that attempt, our improvement requires neither any threshold value nor modification of the stop criterion. Moreover, our newly introduced update rule causes attempted updates to be highly effective.

We refer to a node whose all neighbors have the same label as it does as an \textit{interior} node. Nodes that are not interior are called {\it boundary} nodes. A node that would not change its label if it were to attempt an update is referred to as {\it passive}, while the node that is not passive is called {\it active}. Clearly, all interior nodes are passive by definition. On the other hand, a boundary node could be either active or passive, depending on its neighborhood. Hence, each node may be in one of the three states: passive interior, passive boundary or active boundary. In general, the update rule itself defines a natural end of execution condition, which allow the algorithm to finish execution when every node becomes passive, the situation to which we refer as convergence of a network. Moreover, we maintain a list called \textit{active node list} that contains all currently active nodes.  The general outline of the LPA improvement is as follows:
\begin{description}
\item[1)]At time t=0, construct the active node list containing all the nodes.
\item[2)]Randomly pick an active node, say $i$, from the list and attempt to adopt a new label according to the update rule. Since only active nodes are placed initially on the list and they remain on the list as long as they are active, each node selected for an update will change its label during the update. 
\item[3)]
First, check if the updated node became passive and if so, remove it from the list. Next, check all its neighbors for the change of status in the following three steps. (1) If an interior neighbor  became an active boundary node, add it to the active node list. 
(2) Remove any previously active neighbor that became passive from the active node list. 
(3) Add any previously passive boundary neighbor that became active to the active node list.
\item[4)]If the active node list is empty, stop; otherwise, increase time t by one unit and go to step 2.
\end{description}

The complexity of the improved algorithm is unchanged. Initialization of the active node list requires $O(n)$ time.  Randomly selecting one node takes $O(1)$ time. Updating the node $i$ and its neighbors requires $O(d_i)$, where $d_i$ is the degree of node $i$.  By using the active node list, evaluating the convergence of the whole network is easy and takes exactly $O(1)$ by checking if the list is empty.  In our improvement, the number of iterations needed for the algorithm to converge is equal to the total number of effective updates.

To evaluate the efficiency of our improvement, we have run the original LPA and the LPA with our improvement on a wide range of social networks. The two tested algorithms are denoted as org-LPA and speed-up-LPA in Table ~\ref{table:compareT}, respectively. Social networks used throughout the paper include: 1) karate: Zachary's karate club network \cite{Zac77}; 2) football: the schedule of games of US college football teams \cite{GirvanNewman:2002}; 3) lesmis: interactions between major characters in Victor Hugo's novel \textit{Les Misarables} \cite{Knuth1993}; 4) polbooks: books on American politics co-purchased on Amazon.com \cite{Krebs}; 5) netscience: a network of authors publishing articles on network science \cite{datanetscience}; 6) email: a social network in a company implied by the interactions via email \cite{Guimera2003PRE};  7) eva: a network of a US company; 8) PGP: a network of users of the Pretty-Good-Privacy algorithm for secure information interchange \cite{dataPGP};  9) CA-GrQc: a network of researchers in General Relativity publishing in Arxiv on that topic \cite{dataCA-GrQc}. Note that in all these networks, only the largest connected component is used.

We measured the speed in terms of the number of iterations scaled by the network size $n$. We repeated each experiment 100 times and reported the average. As shown in Table ~\ref{table:compareT}, the new framework does not depend on the network size, and for network size up to ten thousands of nodes, the scaled number of iterations remains below 3. The speed of the algorithm improves by a factor of at least 1.5 for all tested networks, but on larger networks, like email, eva, CA-GrQc and PGP, the algorithm is  6  or more times faster than the original LPA.

\section{A Generalized Update Rule}
\label{sec:genupdaterule}

\subsection{Neighborhood Strength Driven LPA}

The propagation of a label is analogous to epidemic, idea, opinion and information spreading in a network.  By assuming that a node always adopts the label of the majority of its neighbors, the LPA ignores any structures existing in this node neighborhood. This makes the algorithm very simple.  However, in reality, a person adopting a new idea, often follows a neighbor who has more connections to other neighbors because this neighbor has higher number of potential sources of information. For the same reason, when a node joining a group (i.e., changing its label to the one shared by this group) may take into account not only how many members are in this particular group (like the original LPA does) but also how well they are connected to other neighbors of the node executing the update label rule. Following this idea, we generalize the update rule of LPA as follows:

$$L(i)=L\left( \operatorname*{arg\,max}_{C_k} \{ S(C_k) \} \right)$$
where $L$ denotes the label of a node or a community. $C_k$ is the sub-community containing a set of nodes connected to node $i$ and sharing among themselves the same label $k$. $S$ is the score function of a sub-community defined as

$$S(C_k)=\sum_{j \in C_k} \{1+c \cdot h_j(i)\}$$

The first term accounts for the direct link from node $j$ to node $i$. The second term represents the new generalized rule of LPA. Here $h_j(i)$ is the number of links from node $j$ to the entire neighborhood of node $i$, excluding node $i$.  $c$ is a weight between 0 and 1, indicating how much we want to weight the impact of node $j$ on other neighbors. When $c=1$, we place the same weight on all the links in the neighborhood; when $c=0$, we discount those links, except the direct connection to node $i$, which reduces the new rule to the original LPA rule. For simplicity, we do not account for links from $j$ to its own community or to other communities.  Note that we still restrict the process to a local neighborhood of a node, and do not consider links that go outside of this local neighborhood.

Each value of $c$ provides some guidance for the label propagation. The ties between labels to choose from in this process, broken randomly, contribute to  the random output of the algorithm. Assigning the weight $c$ is non-trivial, since LPA has a counterintuitive nature, in which the communities are formed around some local minima instead of the globally optimum value.  Hence, the value of $c$ that provides a good balance between converging quickly and not getting trapped in the undesired local minima leads to better results.  From the experiments below, we find that there is a connection between the factor $c$ and the node Clustering Coefficients (CC) \cite{Watts:1998}.

\section{Evaluation of Performance}
To test the performance of the generalized update rule, we incorporate it into the modified LPA framework in section ~\ref{sec:impspeed} to create a new community detection algorithm and apply it to both computer generated networks and real-world social networks. In the experiments, we explore different values of $c$. More specifically, we analyze the quality of community detection for values of $c$ taken from a set $\{0, 0.05, 0.25, 0.65, 0.8, 1\}$.  When c=1, all links are equally important, while for c=0 (i.e., for original LPA), the links to other neighbors, except those to the node under consideration, are completely ignored. Values of $c$ above 0 and below 1 define how much we favor one of these two extremes. 

During the asynchronous updating scheme, in each step, there is choice of the node to which the update rule is to be applied next. In addition, when there is a tie among labels with the highest scores, there is also a choice of the final label assigned to the node executing the update. Communities detected depend on what selections are made for these choices because each selection may trap the solution in a local minimum. Usually, these selections are done randomly, in which case every run may produce different outcomes. Therefore, statistical measures of quality of communities detected, such as the average and the best, are all important metrics of the algorithm performance. 

\begin{figure*}[th]
	 \begin{minipage}[t]{0.45\linewidth}
			\includegraphics[scale=0.47]{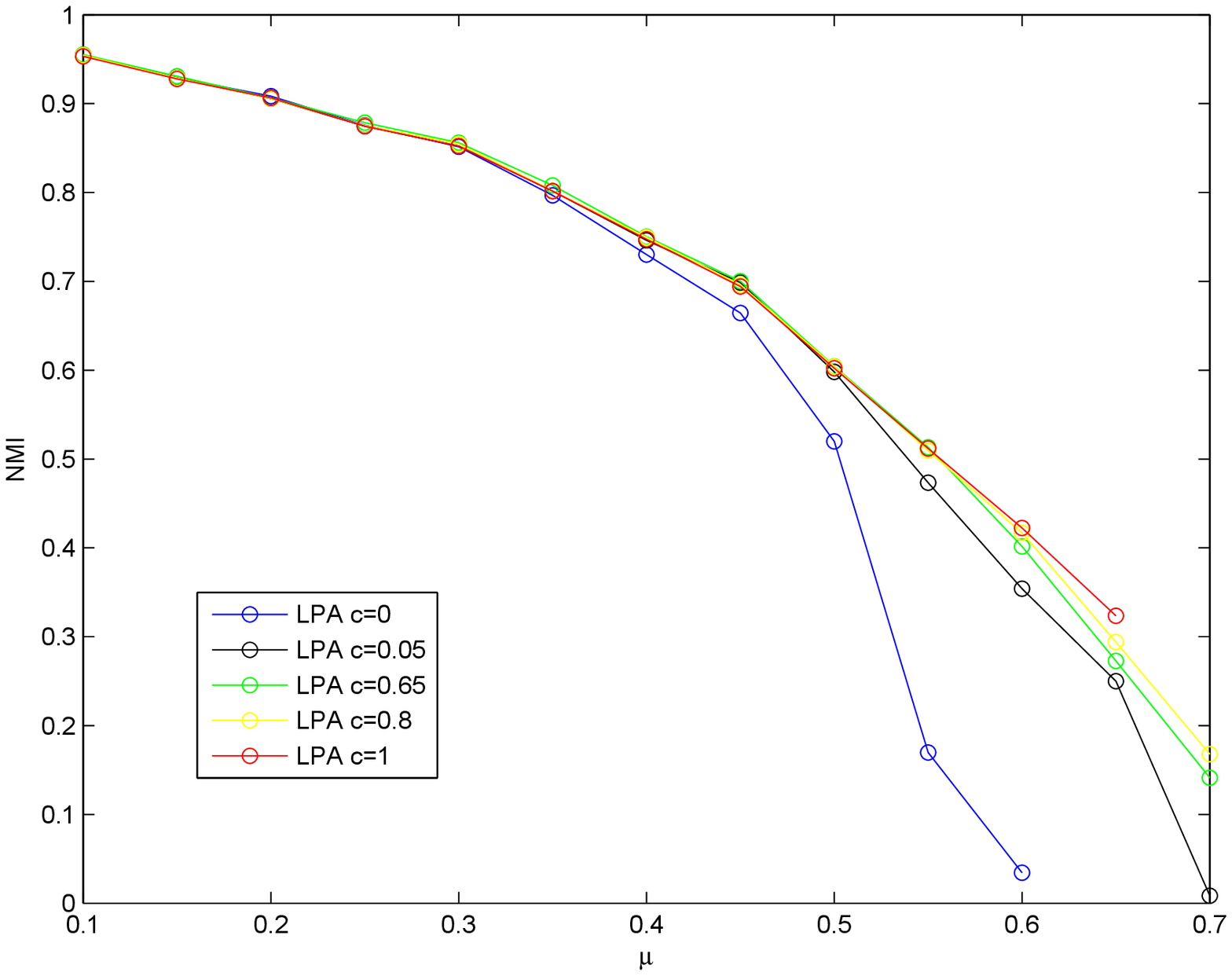}
			\caption{Normalized Mutual Information  of LPA with various $c$'s on LFR networks with $N=1000$ and $<k>=5$.}
			\label{fig:NMI-LPAs}
	\end{minipage}	
	 \hspace{1cm}  
	 \begin{minipage}[t]{0.45\linewidth}
		\includegraphics[scale=0.47]{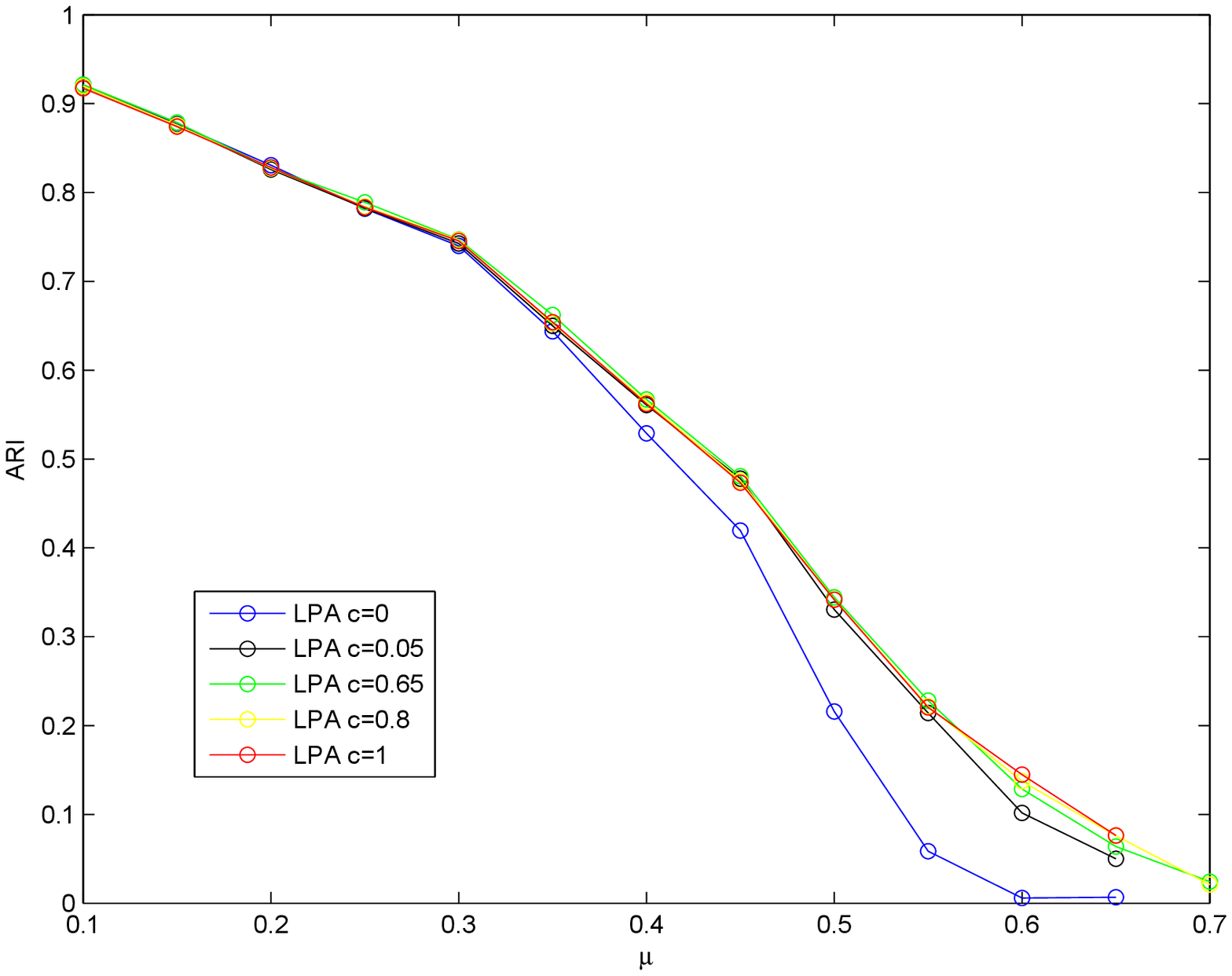}
		\caption{Adjusted Rand Index of LPA with various $c$'s on LFR networks with $N=1000$ and $<k>=5$.}
	 	\label{fig:ARI-LPAs}
	 \end{minipage}
\end{figure*}
\begin{figure*}[th]	
	 \vspace{0.5cm}  
	 \begin{minipage}[t]{0.45\linewidth}
	 \includegraphics[scale=0.47]{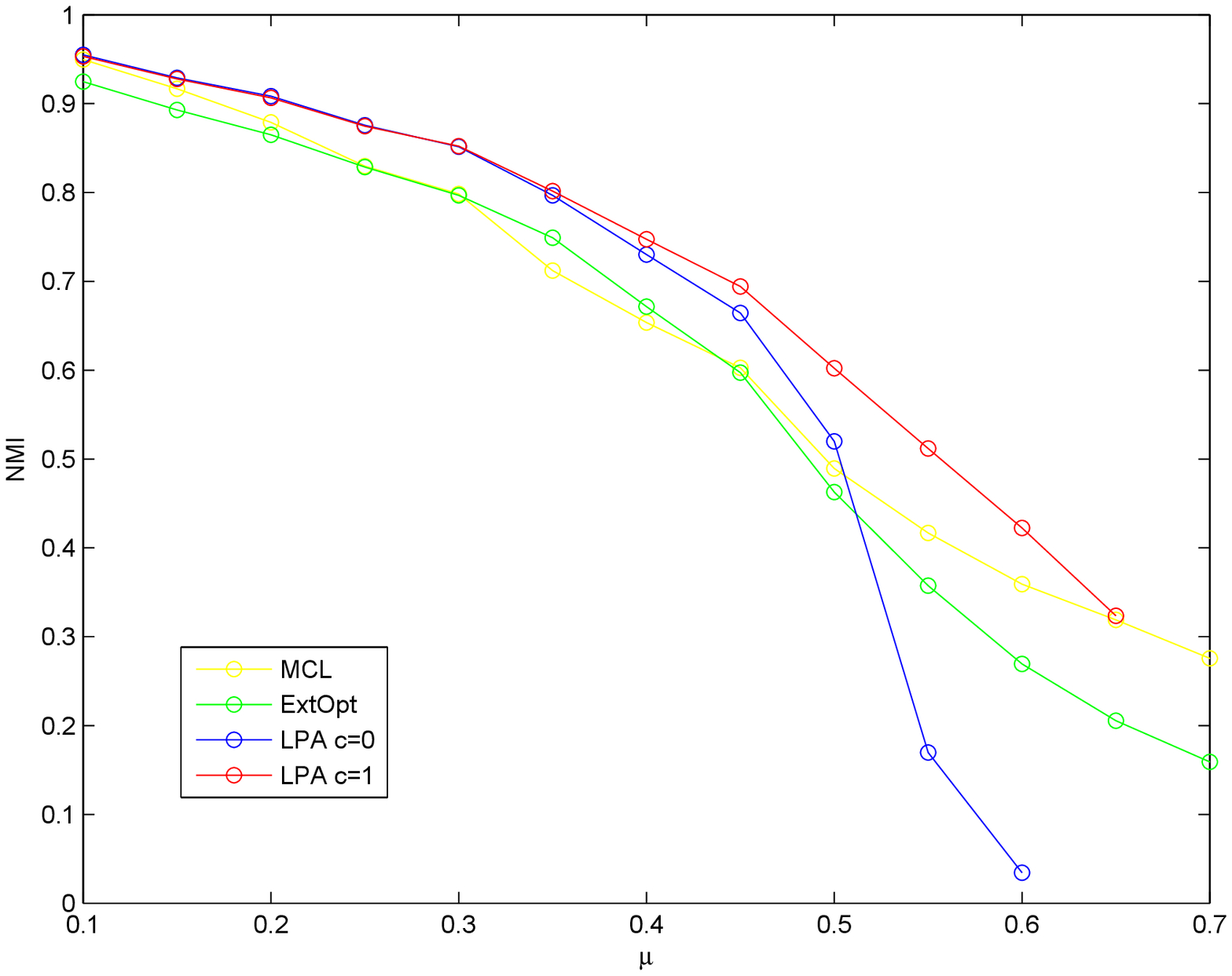}
\caption{Normalized Mutual Information of different detection algorithms on LFR networks with $N=1000$ and $<k>=5$.}
\label{fig:NMI-comp}
	 \end{minipage}
	 \hspace{1cm}  
	 \begin{minipage}[t]{0.45\linewidth}
	 \includegraphics[scale=0.47]{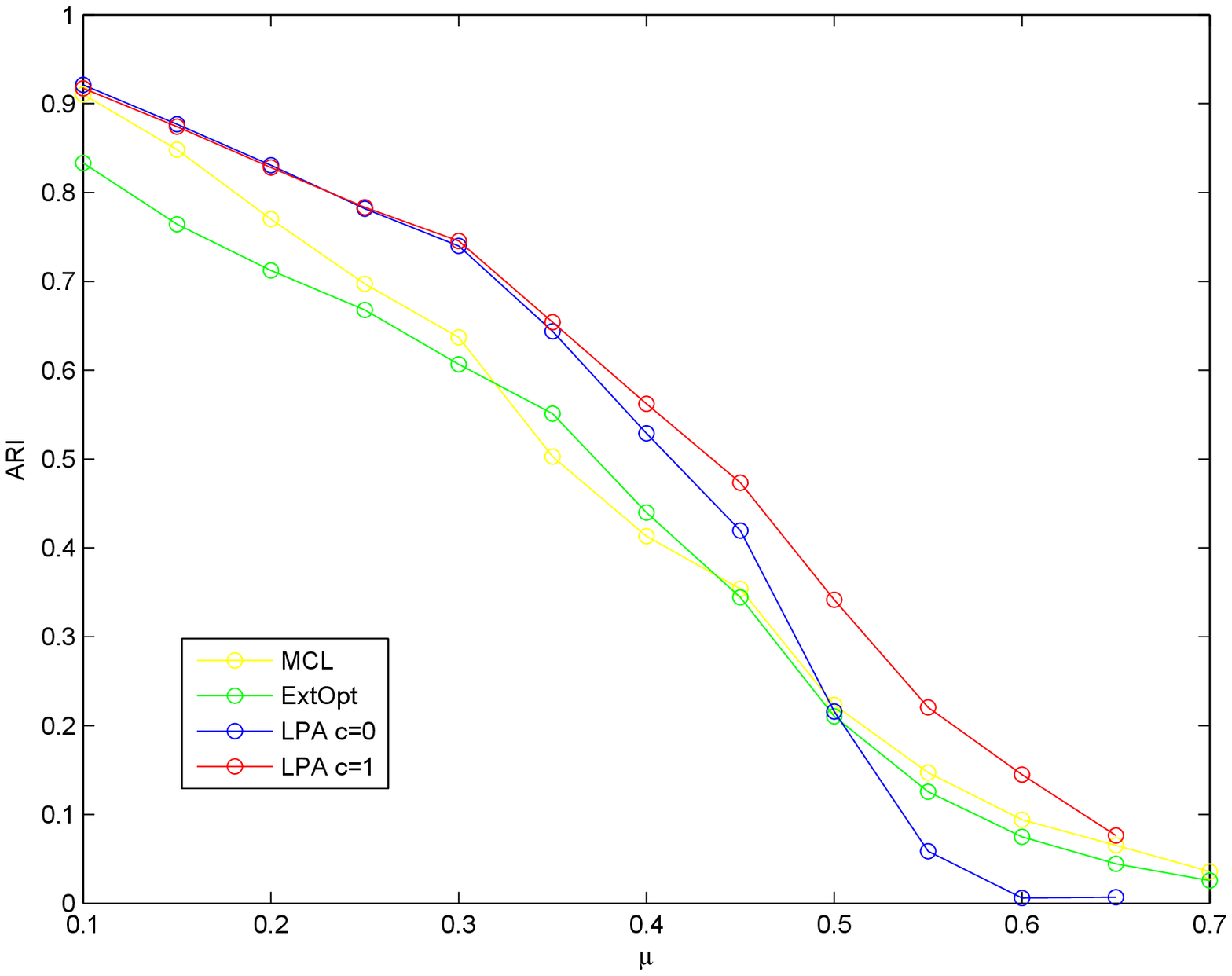}
\caption{Adjusted Rand Index of different detection algorithms on LFR networks with $N=1000$ and $<k>=5$.}
\label{fig:ARI-comp}
	 \end{minipage}
\end{figure*}	 

\subsection{Tests on Computer Generated Networks}

\subsubsection{Benchmark networks and quality measures}
The reason for using computer generated networks is that these networks have well-defined community structures, i.e., we know the pre-assigned true label of each node. We adopted the LFR benchmark \cite{LFR:2008}, which is a special case of the \textit{planted l}-partition model \cite{plantedLpartition:2001}. LFR networks are \textit{similar} to real-world networks because they are characterized, like most of the real-world networks, by heterogeneous distributions of node degrees and community sizes. In our experiments, we used the following fixed parameters: node degrees and community sizes are governed by the power law, with exponents 2 and 1 \cite{plantedLpartition:2001}; the maximum degree is 50; the community sizes vary between 10 and 50; the network size is set at $N=1000$ and the average degree is kept at $<k>=5$. We varied the mixing parameter $\mu$, which is the expected fraction of links of a node connecting to other communities. In other words, each node has $(1-\mu) \cdot <k>$ intra community links on average. The larger the value of $\mu$ is, the weaker the community structure is. 

Many measures have been proposed for quantifying the quality of a partition from a detection algorithm with respect to the known true partition. Each of them has its advantage and disadvantage. In this paper, we carefully have chosen two of them. In \cite{Danon05comparingcommunity}, Danon proposed to use the Normalized Mutual Information (NMI) for network partition, measuring the amount of information correctly extracted by the detection algorithms. NMI is shown to be reliable and is used often in physics literature. The rand index is a measure of the similarity between two partitions, indicating how much they agree in terms of  pairs of nodes. We use its adjusted-for-chance form, namely the Adjusted Rand Index (ARI) \cite{Hubert:1985}. Both NMI and ARI have value 1 for a perfect match and 0 for a random or independent partition \cite{Steinhaeuser:2010}. 

For comparison, two algorithms, extremal optimization (short for ExtOpt) \footnote{\url{http://deim.urv.cat/~aarenas/data}} and MCL \footnote{\url{http://micans.org/mcl/}}, are included in the experiments as references. ExtOpt is a modularity maximization algorithm, which usually obtains high modularity. MCL performs well and is fast in practice. We used 1.4 for the parameter \textit{inflation}, which achieved good results. For our generalized LPA algorithms, we repeated each run 10 times and kept the maximum scores. For each $\mu$, an average values over 10 realizations of networks are reported. 

\subsubsection{Performance analysis}

Fig.~\ref{fig:NMI-LPAs} and Fig.~\ref{fig:ARI-LPAs} demonstrate that the generalized LPA maintains similar behavior when the neighborhood strength is varied (we omitted $c=0.25$ because it is close to $c=0.05$). When $\mu$ is relatively small, the strength ($c>0$)  does not play a significant role, resulting in similar performance for different $c's$. However, when $\mu$ exceeds 0.35, LPA with $c>0$ perform better. Although the average degree is fixed by construction (so is the degree distribution), we observed that the distribution of node clustering coefficients (CC) changes significantly (from $CC=0.54$ for $\mu=0.1$ to $CC=0.25$ for $\mu=0.3$). Since each community is connected in the manner similar to a random graph, increasing $\mu$ leads to smaller average clustering coefficients in the same community. During the updating, the majority rule of LPA becomes weaker. Therefore, adopting the label from a group of neighboring nodes with same label and more \textit{intra} connections reduces the effect of the underlying structure change. This results in a stable performance of LPA with $c>0$ for $0.3< \mu \le 0.5$. Another explanation of the benefit of having $c>0$ is that it tends to maintain the tension between communities during the evolution. Such tension would trap LPA in a sub-optimal solution. However, with $c=0$, LPA loses the tension more quickly. This is especially pronounced for $\mu >0.5$, in which case the algorithm yields a single community most of the time, ending with very low average performance in terms of quality of communities detected.

As shown in Fig.~\ref{fig:NMI-comp} and Fig.~\ref{fig:ARI-comp}, among MCL, ExtOpt and LPA, the LPA outperforms the others consistently for a wide range of  $\mu \le 0.5$ for which community structure is termed strong \cite{radicchi2004defining}, because each node has more links to its own community than to the nodes outside it. In fact, the performances of all algorithms drops sharply before $\mu=0.5$ because the tested networks are sparse which makes detection harder. 
LPA with $c=0$ performs worst beyond $\mu=0.5$. For LPA with $c>0$, the neighbor strength advantage is carried up to $\mu=0.65$, beyond which all versions of LPA  find almost only the trivial solution, i.e., single community. Comparing NMI and ARI measures, it is interesting to observe that ARI is more sensitive to the performance of an algorithm than NMI. For example, sharper change is observed in ARI plots, especially for LPA with $c=0$.


\subsection{Tests on Real-world Social Networks}
As in section ~\ref{sec:impspeed}, we repeated each experiment 100 times. The quality of the detected communities is measured by the modularity Q \cite{newman-2004-69}. In Table ~\ref{table:comparemaxQ}, we separated c=0 (i.e., LPA) and c$>$0 for comparison. 

\subsubsection{Maximum performance}
In Table ~\ref{table:comparemaxQ}, we report the maximum modularity obtained in tests. LPA-Q (c=0) denotes the highest performance for the original LPA and LPA-Q (c$>$0) denotes the highest performance of the algorithm with the new rule (the positive value of $c$ with which that performance was achieved is shown in parenthesis). We also list the modularity either reported in the literature or obtain by other algorithms as a reference.  On karate network, all algorithms with different weights achieve the same maximum modularity (0.416) due to the small size of the network and its limited structure variation. On lesmis network, a c=0.05, slightly divergent from 0, yields higher modularity than LPA.  Football network is a special case, on which the highest modularity (still, by very little margin) is achieved with c=0. For other networks, higher modularity is obtained with c=1.0 (or c=0.25) than with c=0, which shows that the neighbor connections (c$>$0) provide useful information for guiding the evolution of the algorithm.
\begin{table}[hbpt]
\centering
\caption{Evaluation of clustering quality by maximum modularity}
\label{table:comparemaxQ}
\begin{tabular}{|c|c|c|c|c|} \hline
\textbf{Network} & \textbf{n} & \textbf{Q} & \textbf{LPA-Q(c=0)} & \textbf{LPA-Q(c$>$0)}\\ \hline
karate&	34&	0.418 \cite{Duch:2005} &	0.416&	0.416(*.*)\\ \hline
lesmis&	77&	0.540 \cite{PhysRevE.69.026113} &	0.547&	0.550(0.05)\\ \hline
football&	115&	0.601 \cite{GirvanNewman:2002} &	0.604&	0.603(0.05)\\ \hline
polbooks&	105&	0.501 \cite{PhysRevE.70.066111} &	0.499  &	0.526(1.00)\\ \hline
netscience&	379&	0.837 \cite{PhysRevE.70.066111} &	0.816 &	0.825(1.00)\\ \hline
email&	1133&	0.570 \cite{Duch:2005} &	0.535&	0.554(0.25)\\ \hline
eva&	4475&	0.935 \cite{PhysRevE.70.066111} &	0.892&	0.927(1.00)\\ \hline
CA-GrQc&	4730&	0.762 \cite{PhysRevE.70.066111} &	0.760&	0.762(1.00)\\ \hline
PGP&	10680&	0.845 \cite{Duch:2005} &	0.816&	0.840(1.00)\\ \hline
\end{tabular}
\end{table}

\subsubsection{Average performance and stability}
By repeating many runs, one can measure the average performance to evaluate variability of the quality of communities detected in those runs. Table ~\ref{table:compareavgQ} shows the results for networks with at least 100 nodes.  In our experiments, we observed that some LAP runs reached complete consensus (i.e., the result is a single community, in which Q=0). For algorithms with c$>$0, this rarely happens. We remove such cases when we compute the average modularity for LPA.  As shown, algorithm with c$>$0 obtains higher average performance than LPA with c=0 on most networks. A trend that favors c$>$0 is clearly shown, except for football network. In other words, as $c$ increases (more weight for neighbor connections), the algorithm becomes more stable on most of these networks. One observation from larger networks (n$>$1000) is that a $c$ that achieves a higher maximum modularity also yields a better average performance.  
\begin{table}[hbpt]
\centering
\caption{The average modularity of LPA with various \textit{c's}}
\label{table:compareavgQ}
\begin{tabular}{|c|c|c|c|c|c|c|} \hline
\textbf{Network} & \textbf{c=1.0}  & \textbf{c=0.8}  & \textbf{c=0.65}  & \textbf{c=0.25}  & \textbf{c=0.05}  & \textbf{LPA(c=0)} \\ \hline
football&	0.567&	0.568&	0.568&	0.576&	0.577&	0.590\\ \hline
polbooks&	0.521&	0.521&	0.519&	0.509&	0.507&	0.487\\ \hline
netscience&	0.804&	0.804&	0.804&	0.803&	0.802&	0.798\\ \hline
email&	0.490&	0.485&	0.471&	0.408&	0.298&	0.230\\ \hline
eva&	0.919&	0.916&	0.911&	0.891&	0.890&	0.890\\ \hline
CA-GrQc&	0.756&	0.753&	0.753&	0.750&	0.748&	0.752\\ \hline
PGP&	0.830&	0.824&	0.822&	0.807&	0.802&	0.802\\ \hline
\end{tabular}
\end{table}

\subsubsection{The weight factor $c$ and the clustering coefficient distribution}

\begin{figure}[!t]
\centering
\includegraphics[scale=0.47]{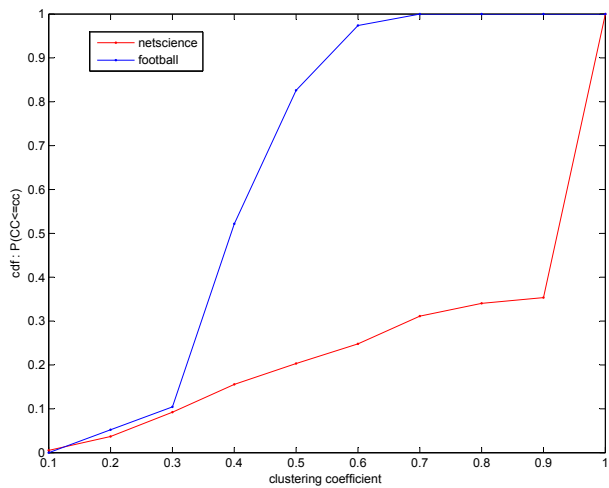}
\caption{The clustering coefficients distribution for the football network with average cc=0.4032 and for netscience network with average cc=0.8125.}
\label{fig:cc-distribution}
\end{figure}

Both Table ~\ref{table:comparemaxQ} and Table ~\ref{table:compareavgQ} show that for many networks, c$>$0 (often c=1) yields better performance than c=0. Our conjecture is that $c$ is strongly related to both degree distribution and clustering coefficient distribution. Given that all tested networks are real-world networks, their degree distributions are similar. Hence, we discuss here the clustering coefficient distribution.  In Fig.~\ref{fig:cc-distribution}, we show the cumulative probability distribution of clustering coefficient (abbreviated as cc), i.e., $P(CC \leq cc)$, where $cc=z/(d_i(d_i-1)/2)$ for a node with $z$ links in the neighborhood (for $d_i=1$, $cc=1$). All cc's are clustered in bins with bin width 0.1. In the case where the algorithm favors smaller $c$, e.g., football network, we observe a distribution shown in Fig. ~\ref{fig:cc-distribution} (blue). If we consider nodes with cc$>$0.9 as \textit{highly clustered}, then in football network, most nodes are not strongly clustered (with cc$<$0.6). Another distribution in Fig. ~\ref{fig:cc-distribution} (red) presents a different feature, that is, nodes with smaller cc are roughly uniformly distributed, and there is a large fraction of nodes with high cc. For example, in netscience network (a co-authorship network), there are many small communities with either a key node (a core researcher) connected either to many isolated nodes or a strong inter-connected group (e.g., a research lab), both of which account for large values of cc for the corresponding nodes. Since the new rule with positive $c$ tends to prefer a sub-community with more connections inside the neighborhood (in some sense it implies that such sub-community also has stronger interconnections), it is consistent with the feature observed. Therefore, it is not surprising that the new rule works more efficiently than LPA (with c=0) on networks like netsience, email, eva, CA-GrQc and PGP that all have similar shape of clustering coefficient distribution.  

\begin{figure*}[th]		 
	 \begin{minipage}[t]{0.45\linewidth}
		\includegraphics[scale=0.47]{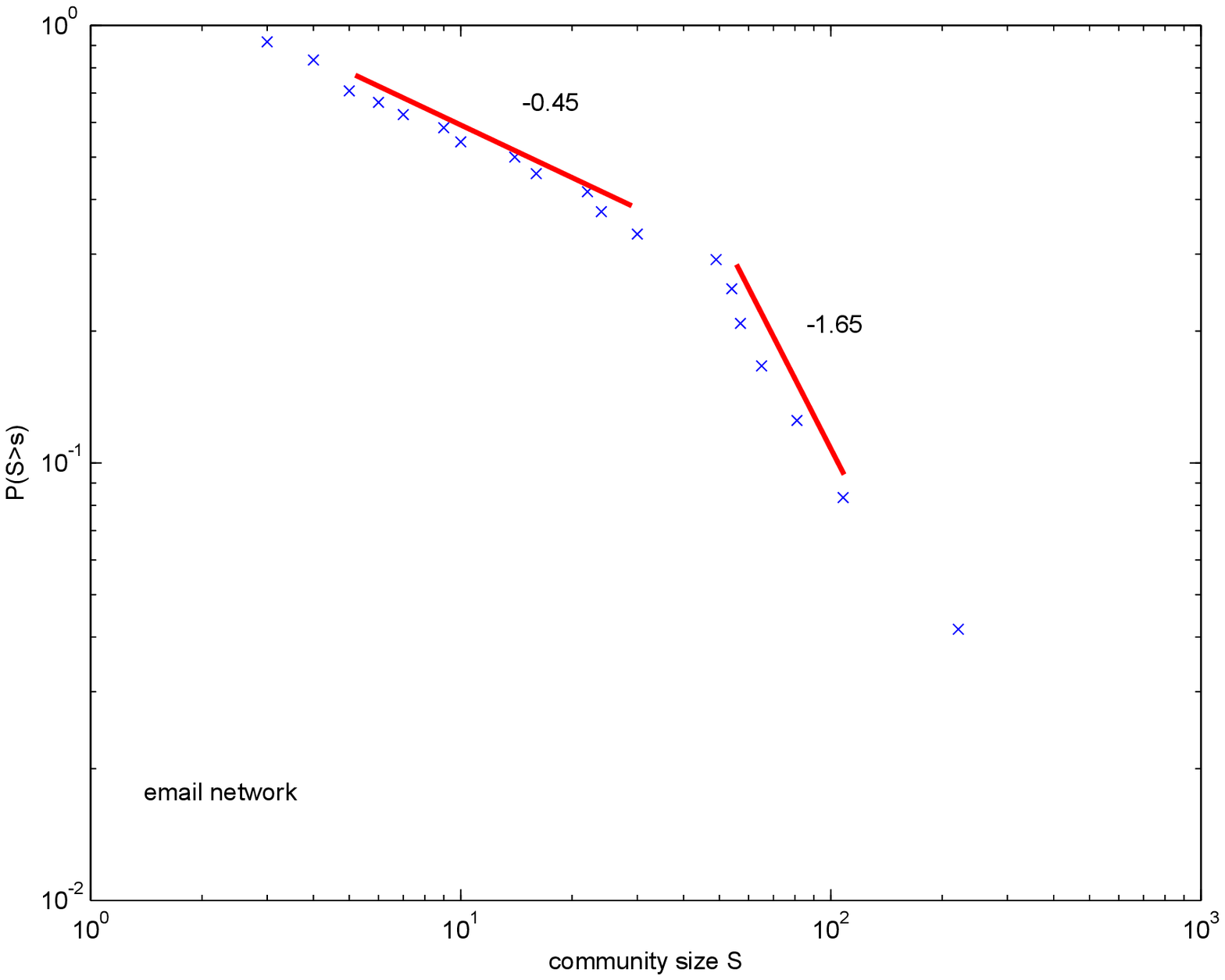}
		\caption{The two-part power law community size distribution for the email network found with c=0.25. The results are taken from a specific run with highest modularity 0.554.}
		\label{fig:email-comdist}
	 \end{minipage}
	 \hspace{1cm}  
	 \begin{minipage}[t]{0.45\linewidth}
		\includegraphics[scale=0.47]{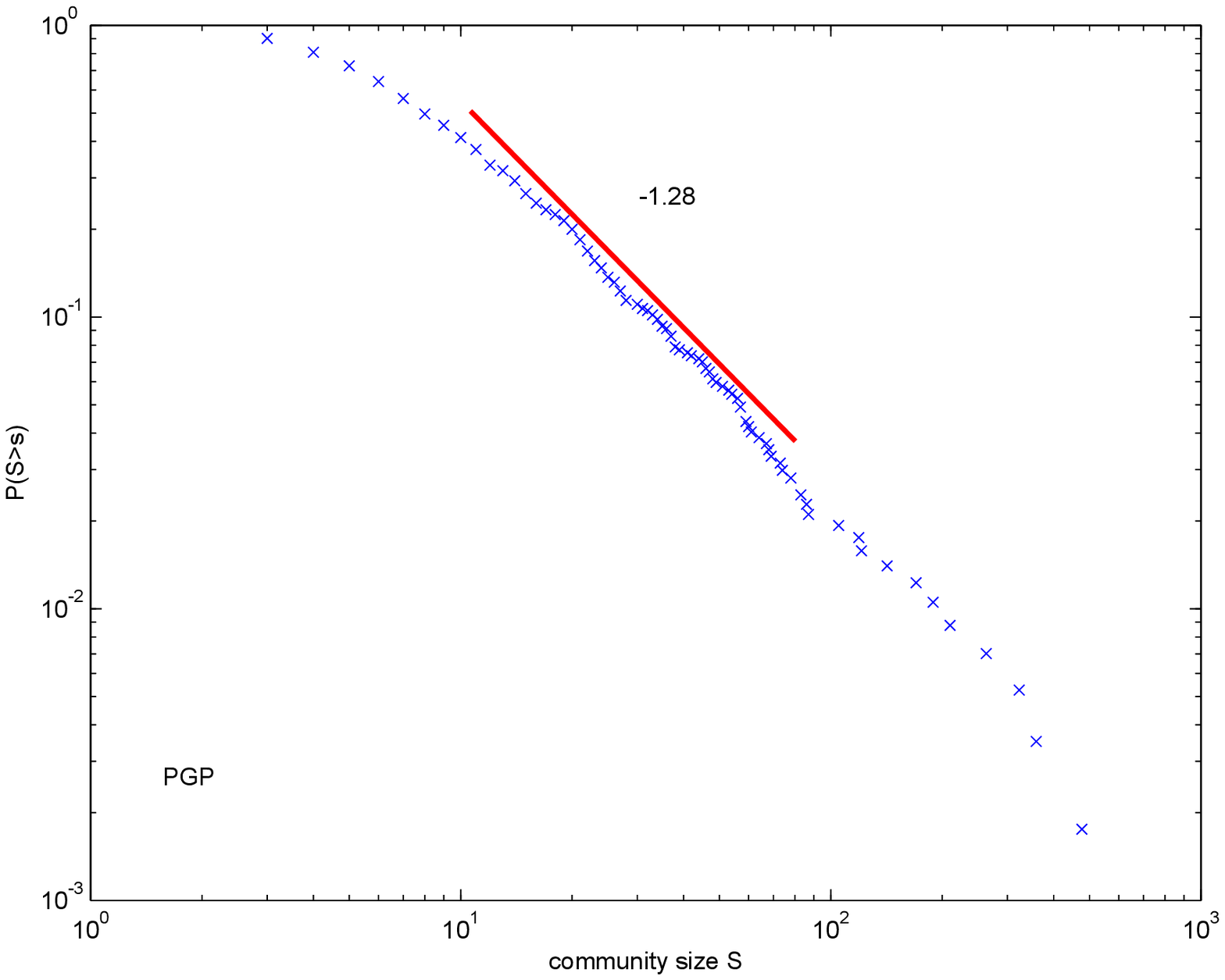}
		\caption{The power law community size distribution for the PGP network found with c=1.00.  The results are taken from a specific run with highest modularity 0.840.}
		\label{fig:pgp-comdist}
	 \end{minipage}
\end{figure*}	 

\subsubsection{Community size}
Like in the case of the LPA algorithm, the community size found by the algorithm with c$>$0 follows a power law distribution, $P(S>s) \propto s^\alpha $. The exponent $\alpha$ estimated by the clustering result with the maximum modularity for the PGP network (see Fig. ~\ref{fig:pgp-comdist} with c=1.0) is about -1.28. For the two-part power law for the email network (see Fig. ~\ref{fig:email-comdist} with c=0.25), two values of $\alpha$ are -1.65 and -0.45. This is consistent with previous observations \cite{Raghavan:2007} discussed in \cite{PhysRevE.70.066111, newman-2004-69}.

\section{Conclusions}
\label{conclusion}
In this paper, we presented a new community detection algorithm that improves both the speed and quality of detected communities when compared to the original LPA algorithm. The generalized update rule allows us to incorporate useful neighborhood information.  Both maximum and average detected community quality improves for most of the tested networks. The parameter $c$ is related to an interesting feature of the networks, i.e., the clustering coefficient distribution, that explains the difference in optimal value of $c$ for many different networks. However, the selection of $c$ is not yet fully understood, and therefore it is the subject worthy of further study.  Extending our approach to overlapping and online community detection is another future research direction that we plan to pursue.

\section*{Acknowledgment}

This work was supported in part by the Army Research
Laboratory under Cooperative Agreement Number
W911NF-09-2-0053 and by the Office of Naval Research
Grant No. N00014-09-1-0607. The views and
conclusions contained in this document are those of the
authors and should not be interpreted as representing
the official policies either expressed or implied of the
Army Research Laboratory, the Office of Naval Research, or the U.S. Government.

\bibliographystyle{IEEEtran}  
\bibliography{IEEEabrv,CommunityBIB}

\begin{thebibliography}{10}
\providecommand{\url}[1]{#1}
\csname url@samestyle\endcsname
\providecommand{\newblock}{\relax}
\providecommand{\bibinfo}[2]{#2}
\providecommand{\BIBentrySTDinterwordspacing}{\spaceskip=0pt\relax}
\providecommand{\BIBentryALTinterwordstretchfactor}{4}
\providecommand{\BIBentryALTinterwordspacing}{\spaceskip=\fontdimen2\font plus
\BIBentryALTinterwordstretchfactor\fontdimen3\font minus
  \fontdimen4\font\relax}
\providecommand{\BIBforeignlanguage}[2]{{%
\expandafter\ifx\csname l@#1\endcsname\relax
\typeout{** WARNING: IEEEtran.bst: No hyphenation pattern has been}%
\typeout{** loaded for the language `#1'. Using the pattern for}%
\typeout{** the default language instead.}%
\else
\language=\csname l@#1\endcsname
\fi
#2}}
\providecommand{\BIBdecl}{\relax}
\BIBdecl

\bibitem{Hoerdt03completenessof}
M.~Hoerdt and U.~Louis, ``Completeness of the internet core topology collected
  by a fast mapping software,'' in \emph{Proc of the 11th International
  Conference on Software, Telecommunications and Computer Networks}, 2003, pp.
  257--261.

\bibitem{Broder:2000:GSW:346241.346290}
A.~Broder, R.~Kumar, F.~Maghoul, P.~Raghavan, S.~Rajagopalan, R.~Stata,
  A.~Tomkins, and J.~Wiener, ``Graph structure in the web,'' in \emph{Proc of
  the Ninth International Conference on the World Wide Web}, 2003, pp. 15--19.

\bibitem{citeulike:128}
M.~E.~J. Newman, ``The structure of scientific collaboration networks,''
  \emph{Proc Natl Acad Sci USA}, vol.~98, no.~2, pp. 404--409, January 2001.

\bibitem{RefWorks:380}
J.~Scott, \emph{Social Network Analysis: A Handbook}.\hskip 1em plus 0.5em
  minus 0.4em\relax Sage, 2000.

\bibitem{Fell:2000}
D.~A. Fell and A.~Wagner, ``The small world of metabolism,'' \emph{Nature
  Biotechnology}, vol.~18, no.~11, pp. 1121--1122, November 2000.

\bibitem{Danon05comparingcommunity}
L.~Danon, J.~Duch, A.~Arenas, and A.~Diaz-guilera, ``Comparing community
  structure identification,'' \emph{Journal of Statistical Mechanics: Theory
  and Experiment}, vol. 9008, p. 09008, 2005.

\bibitem{GirvanNewman:2002}
M.~Girvan and M.~E.~J. Newman, ``Community structure in social and biological
  networks,'' \emph{Proc Natl Acad Sci USA}, vol.~99, no.~12, pp. 7821--7826,
  Jun. 2002.

\bibitem{newman-2004-69}
M.~E.~J. Newman, ``Fast algorithm for detecting community structure in
  networks,'' \emph{Phys. Rev. E}, vol.~69, p. 066133, 2004.

\bibitem{PhysRevE.69.026113}
M.~E.~J. Newman and M.~Girvan, ``Finding and evaluating community structure in
  networks,'' \emph{Phys. Rev. E}, vol.~69, p. 026113, 2004.

\bibitem{PhysRevE.70.066111}
A.~Clauset, M.~E.~J. Newman, and C.~Moore, ``Finding community structure in
  very large networks,'' \emph{Phys. Rev. E}, vol.~70, p. 066111, Dec 2004.

\bibitem{springerlink:10.1140/epjb/e2004-00125-x}
F.~Wu and B.~Huberman, ``Finding communities in linear time: A physics
  approach,'' \emph{The European Physical Journal B - Condensed Matter and
  Complex Systems}, vol.~38, pp. 331--338, 2004.

\bibitem{PhysRevE.74.036104}
M.~E.~J. Newman, ``Finding community structure in networks using the
  eigenvectors of matrices,'' \emph{Phys. Rev. E}, vol.~74, p. 036104, Sep
  2006.

\bibitem{Duch:2005}
J.~Duch and A.~Arenas, ``Community detection in complex networks using extremal
  optimization,'' \emph{Phys. Rev. E}, vol.~72, p. 027104, 2005.

\bibitem{Raghavan:2007}
U.~N. Raghavan, R.~Albert, and S.~Kumara, ``Near linear time algorithm to
  detect community structures in large-scale networks,'' \emph{Phys. Rev. E},
  vol.~76, p. 036106, 2007.

\bibitem{GirNew02}
M.~Girvan and M.~E.~J. Newman, ``Community structure in social and biological
  networks,'' \emph{Proc Natl Acad Sci USA}, vol.~99, no.~12, pp. 7821--7826,
  June 2002.

\bibitem{WakitaTsurumi:2007}
K.~Wakita and T.~Tsurumi, ``Finding community structure in mega-scale social
  networks,'' in \emph{Proc of the 16th international conference on World Wide
  Web}, 2007, pp. 1275--1276.

\bibitem{SchuetzCaflisch:2008}
P.~Schuetz and A.~Caflisch, ``Efficient modularity optimization by multistep
  greedy algorithm and vertex mover refinement,'' \emph{Phys. Rev. E}, vol.~77,
  p. 046112, 2008.

\bibitem{Blondel-2008}
V.~D. Blondel, J.-L. Guillaume, R.~Lambiotte, and E.~Lefebvre, ``Fast unfolding
  of communities in large networks,'' \emph{J. Stat. Mech.}, October 2008.

\bibitem{ZhouLipowsky:2006}
H.~Zhou and R.~Lipowsky, ``Network brownian motion: A new method to measure
  vertex-vertex proximity and to identify communities and subcommunities,''
  \emph{Lect. Notes Comput. Sci.}, vol. 3038, no. 1062-1069, 2004.

\bibitem{Hu:2008}
Y.~Hu, M.~Li, P.~Zhang, Y.~Fan, and Z.~Di, ``Community detection by signaling
  on complex networks,'' \emph{Phys. Rev. E}, vol. 78(1), p. 016115, 2008.

\bibitem{DBLP:journals/jgaa/PonsL06}
P.~Pons and M.~Latapy, ``Computing communities in large networks using random
  walks,'' \emph{J. Graph Algorithms Appl.}, vol.~10, no.~2, pp. 191--218,
  2006.

\bibitem{StijnDongen:2000}
S.~van Dongen, ``Graph clustering by flow simulation,'' Ph.D. dissertation,
  University of Utrecht, 2000.

\bibitem{ShiMalik:2007}
J.~Shi and J.~Malik, ``Normalized cuts and image segmentation,'' \emph{IEEE
  Trans. Pattern Anal. Mach. Intell.}, vol.~22, no.~8, p. 888¨C901, 2000.

\bibitem{WhiteSmyth:2005}
S.~White and P.~Smyth, ``A spectral clustering approach to finding communities
  in graphs,'' in \emph{Proc of SIAM International Conference on Data Mining},
  2005, pp. 76--84.

\bibitem{Capocci:2005}
A.~Capocci, V.~Servedio, G.~Caldarelli, and F.~Colaiori, ``Detecting
  communities in large networks,'' \emph{Physica A}, vol. 352, pp. 669--676,
  2005.

\bibitem{Blatt:1996}
M.~Blatt, S.~Wiseman, and E.~Domany, ``Superparamagnetic clustering of data,''
  \emph{Phys. Rev. Lett.}, vol.~76, pp. 3251--3254, 1996.

\bibitem{Reichardt:2004}
J.~Reichardt and S.~Bornholdt, ``Detecting fuzzy community structures in
  complex networks with a potts model,'' \emph{Phys. Rev. Lett.}, vol.~93, p.
  218701, 2004.

\bibitem{randomfield:2006}
S.-W. Son, H.~Jeong, and J.-D. Noh, ``Random field ising model and community
  structure in complex networks,'' \emph{Eur. Phys. J. B}, vol.~50, p. 431,
  2006.

\bibitem{Fortunato:2007}
S.~Fortunato and M.~Barthelemy, ``Resolution limit in community detection,''
  \emph{Proc. Natl. Acad. Sci. USA}, vol. 104, pp. 36--41, 2007.

\bibitem{Kumpula:2007}
J.~Kumpula, J.~Saramaki, K.~Kaski, and J.~Kertesz, ``Limited resolution in
  complex network community detection with potts model approach,'' \emph{Eur.
  Phys. J. B}, vol.~56, pp. 41--45, 2007.

\bibitem{Bagrow:2005}
J.~Bagrow and E.~Bollt, ``A local method for detecting communities,''
  \emph{Phys. Rev. E}, vol.~72, p. 046108, 2005.

\bibitem{Costa:2004}
L.~da~Fontoura~Costa, ``Hub-based community finding,'' \emph{condmat/0405022},
  2004.

\bibitem{WuHuberman:2004}
F.~Wu and B.~A. Huberman, ``Finding communities in linear time: a physics
  approach,'' \emph{Eur. Phys. J. B}, vol.~38, p. 331, 2004.

\bibitem{TibKert:2008}
G.~Tib\'ely and J.~Kert\'esz, ``On the equivalence of the label propagation
  method of community detection and a potts model approach,'' \emph{Physica A},
  vol. 387, pp. 4982--4984, 2008.

\bibitem{Barber:2009}
M.~J. Barber, ``Detecting network communities by propagating labels under
  constraints,'' \emph{Phys. Rev. E}, vol.~80, p. 026129, 2009.

\bibitem{Leung:2009}
I.~X.~Y. Leung, P.~Hui, P.~Li¨°, and J.~Crowcroft, ``Towards real-time
  community detection in large networks,'' \emph{Phys. Rev. E}, vol.~79, p.
  066107, 2009.

\bibitem{Gregory:2010}
S.~Gregory, ``Finding overlapping communities in networks by label
  propagation,'' \emph{New J. Phys.}, vol.~12, p. 103018, 2010.

\bibitem{Zac77}
W.~Zachary, ``An information flow model for conflict and fission in small
  groups,'' \emph{Journal of Anthropological Research}, vol.~33, pp. 452--473,
  1977.

\bibitem{Knuth1993}
D.~E. Knuth, \emph{The Stanford GraphBase: A Platform for Combinatorial
  Computing}.\hskip 1em plus 0.5em minus 0.4em\relax Addison-Wesley, 1993.

\bibitem{Krebs}
V.~Krebs, ``http://www.orgnet.com/ (unpublished).''

\bibitem{datanetscience}
M.~E.~J. Newman, ``The structure and function of complex networks,'' \emph{SIAM
  Review}, vol.~45, pp. 167--256, 20200307.

\bibitem{Guimera2003PRE}
R.~Guimer\`a, L.~Danon, A.~D\'\i{}az-Guilera, F.~Giralt, and A.~Arenas,
  ``Self-similar community structure in a network of human interactions,''
  \emph{Phys. Rev. E}, vol.~68, p. 065103, Dec 2003.

\bibitem{dataPGP}
M.~Boguna, R.~Pastor-Satorras, A.~Diaz-Guilera, and A.~Arenas, ``Models of
  social networks based on social distance attachment,'' \emph{Physical Review
  E}, vol.~70, p. 056122, 2004.

\bibitem{dataCA-GrQc}
J.~Leskovec, J.~Kleinberg, and C.~Faloutsos, ``Graph evolution: Densification
  and shrinking diameters,'' \emph{ACM Transactions on Knowledge Discovery from
  Data}, vol.~1, 2007.

\bibitem{Watts:1998}
D.~J. Watts and S.~Strogatz, ``Collective dynamics of small-world networks,''
  \emph{Nature}, vol. 393, p. 440¨C442, 1998.

\bibitem{LFR:2008}
A.~Lancichinetti, S.~Fortunato, and F.~Radicchi, ``Benchmark graphs for testing
  community detection algorithms,'' \emph{Phys. Rev. E}, vol.~78, p. 046110,
  2008.

\bibitem{plantedLpartition:2001}
A.~Condon and R.~M. Karp, ``Algorithms for graph partitioning on the planted
  bisection model,'' \emph{Random Structures and Algorithms}, vol.~18, pp.
  116--140, 2001.

\bibitem{Hubert:1985}
L.~Hubert and P.~Arabie, ``Comparing partitions,'' \emph{Journal of
  Classification}, vol.~2, pp. 193--218, 1985.

\bibitem{Steinhaeuser:2010}
K.~Steinhaeuser and N.~V. Chawla, ``Identifying and evaluating community
  structure in complex networks,'' \emph{Pattern Recognition Letters}, vol.~31,
  pp. 413--421, 2010.

\bibitem{radicchi2004defining}
F.~Radicchi, C.~Castellano, F.~Cecconi, V.~Loreto, and D.~Parisi, ``Defining
  and identifying communities in networks,'' \emph{Proc Natl Acad Sci USA},
  vol. 101, no.~9, p. 2658, 2004.

\end{thebibliography}

\end{document}